\documentclass[prl,twocolumn,superscriptaddress,floats]{revtex4}

\usepackage[dvips]{graphicx}
\usepackage{amsmath}
\usepackage{booktabs}
\usepackage{dcolumn}

\usepackage{rotating}

\newlength{\La} 

\newlength{\Lb} 

\newlength{\Lc} 

\newcolumntype{d}{D{.}{.}{-1}}

\newcommand{\tzg}{t$_{2g}$}

\newcommand{\srruo}{Sr$_2$RuO$_4$}

\newcommand{\gbd}{$\gamma$-band}

\newcommand{\gsh}{$\gamma$-sheet}
\newcommand{\ash}{$\alpha$-sheet}
\newcommand{\bsh}{$\beta$-sheet}

\newcommand {\qi}{$\mathbf{q_i}$}
\newcommand {\vQ}{$\mathbf{Q}$}
\newcommand {\vq}{$\mathbf{q}$}

\newcommand {\om}{$\omega$}

\newlength{\figwidth}

\divide\figwidth by 2

\begin{document}

\advance\vsize by 2 cm

\title{Anisotropy of the incommensurate fluctuations in \srruo ~:
a study with polarized neutrons }

\author{M.~ Braden$^*$ }
\affiliation{II. Physikalisches Institut, Universit\"at zu K\"oln,
Z\"ulpicher Str. 77, D-50937 K\"oln, Germany}

\author{P. Steffens}
\affiliation{II. Physikalisches Institut, Universit\"at zu K\"oln,
Z\"ulpicher Str. 77, D-50937 K\"oln, Germany}

\author{Y. Sidis}
\affiliation{Laboratoire L\'eon Brillouin, C.E.A./C.N.R.S.,
F-91191 Gif-sur-Yvette CEDEX, France}

\author{J. Kulda}
\affiliation{Institut Laue-Langevin, Bo\^ite Postale 156, 38042
Grenoble Cedex 9, France}

\author{P. Bourges}
\affiliation{Laboratoire L\'eon Brillouin, C.E.A./C.N.R.S.,
F-91191 Gif-sur-Yvette CEDEX, France}

\author{S. Hayden}
\affiliation{H. H. Wills Physics Laboratory, University of
Bristol, United Kingdom}

\author{N. Kikugawa}
\affiliation{Department of Physics, Kyoto University, Kyoto
606-8502, Japan}

\author{Y. Maeno}
\affiliation{ International Innovation Center and Department of
Physics, Kyoto , 606-8502 Kyoto, Japan}

\date{\today, \textbf{preprint}}


\begin{abstract}

The anisotropy of the magnetic incommensurate fluctuations in
\srruo ~has been studied by inelastic neutron scattering with
polarized neutrons. We find a sizeable enhancement of the out of
plane component by a factor of two for intermediate energy
transfer which appears to decrease for higher energies. Our
results qualitatively confirm calculations of the spin-orbit
coupling, but the experimental anisotropy and its energy
dependence are weaker than predicted.

\end{abstract}

\maketitle

The unconventional superconductor \srruo ~ has been studied in
detail \cite{maeno1,maeno-rev}, but the exact pairing mechanism
remains still unclear. Inspired by the ferromagnetic order in the
perovskite SrRuO$_3$ a p-wave triplet pairing mediated through
ferromagnetic fluctuations was suggested shortly after the
discovery of the superconductivity \cite{rice-sigrist}. In the
meanwhile, there is experimental evidence that the order
parameter indeed has a triplet character, but the ferromagnetic
fluctuations have not yet been detected directly\cite{maeno-rev}.
The Fermi-surface of \srruo ~ has been analyzed precisely
\cite{maeno-rev,mazin1,mazin2}: it consists of three sheets
related to the different \tzg -orbitals. States with
$d_{xy}$-character form a two-dimensional band (\gbd ) with a
cylindrical Fermi-surface sheet (\gsh ).
The states with $d_{xz}$- or $d_{yz}$-character
form nearly one-dimensional bands with flat Fermi-surface sheets
(\ash ~and \bsh ). Due to hybridization effects these flat sheets
get rounded but, still, there is pronounced nesting, as it has
been first calculated in LDA \cite{mazin2}.

An inelastic neutron scattering (INS) experiment determines the
imaginary part of the generalized susceptibility directly
\cite{Lovesey};

\begin{equation}
\frac{d^2 \sigma}{d \Omega d \omega}= \frac{k_f}{k_i} \frac{r_0^2
F^2({\bf Q})}{\pi (g\mu _B)^2} \sum_{\alpha=x,y,z}
(1-\frac{Q_{\alpha}^2}{|{\bf Q}|^2})
\frac{\chi_{\alpha\alpha}"({\bf Q}, \omega)}{1-\exp(-\hbar \omega
/ k_BT)}\; \label{eq:INS}
\end{equation}
where k$_i$ and k$_f$ are the incident and final neutron wave
vectors, $r_0^2$=0.292 barn, $F({\bf Q})$ is the magnetic form
factor. (The scattering-vector ${\bf Q}$ can be split into ${\bf
Q}$ =${\bf q}$+${\bf G}$, where ${\bf q}$ lies in the first
Brillouin-zone and ${\bf G}$ is a zone-center. All reciprocal
space coordinates $(Q_x,Q_y,Q_l)$ are given in reduced lattice
units of 2$\pi$/a or 2$\pi$/c.) The first INS experiment
\cite{sidis} confirmed the nesting features arising from the
one-dimensional bands but was unable to detect any
quasi-ferromagnetic signal. The dynamic susceptibility at
moderate energies is dominated by the incommensurate fluctuations
occurring very close to the calculated position at ${\bf
q_i}=(0.30,0.30,q_z)$. Since then, several theoretical approaches
explored the possible role of the different bands in the
superconducting pairing
\cite{sato,kuwabara,takimoto,nomura,kuroki}.

From the integration over the Fermi-surface, Mazin and Singh
\cite{mazin2} conclude that the incommensurate fluctuations favor
a d-wave singlet pairing in contrast to the experimental
observations. However, such analysis completely neglects the
anisotropy of the susceptibility
$\chi"_{a,b}:=\chi"_{xx}=\chi"_{yy}\not= \chi"_{zz}$. More recent
analyses conclude that the incommensurate fluctuations indeed may
lead to triplet pairing provided that there is a strong
anisotropy with $\chi"_{zz}(\bf q_i)$ much larger than
$\chi"_{a,b}(\bf q_i)$\cite{sato,kuwabara,kuroki}. The question
about the anisotropy of the incommensurate fluctuations is hence
essential for the superconducting pairing.

Ng and Sigrist analyzed the Lindhard susceptibility taking the
spin orbit coupling into account and indeed find an enhancement of
the out of plane component close to the incommensurate position
\qi ~\cite{ng}. This anisotropy appears to be rather modest in the
bare susceptibility $\chi ^0 ({\bf q})$ but may become essential
in the interaction enhanced susceptibility, which in the most
simple RPA treatment is given by :

\begin{equation}
\chi_{\alpha \alpha} ({\bf q})= {{\chi_{\alpha \alpha} ^0 ({\bf
q})}\over{1-I({\bf q}) \cdot \chi_{\alpha \alpha}^0 ({\bf q})}}
,\label{eq:chi}
\end{equation}
where $\alpha$ labels the components and $I(q)$ is the interaction
parameter, which may be anisotropic.  Since the enhancement
parameter $S({\bf q})=I({\bf q})\chi^0({\bf q})$ is close to one
in \srruo ~ near \qi , the enhanced susceptibility $\chi ({\bf
q},\omega)$ is very sensitive to any, even small change in
$\chi^0$. For the resulting $\chi "({\bf q_i}, 6{\rm meV})$,
Eremin et al. obtain an anisotropy of at least one order of
magnitude by full analysis of the spin orbit coupling
\cite{eremin}. On the experimental side, evidence for an
anisotropy was found in the NMR $1\over T_1T$ measurements by
Ishida et al. who obtain a factor of three for the
$\chi"_{zz}\over \chi"_{a,b}$ ratio at the NMR frequency
\cite{ishida2001}. Qualitatively the anisotropy is also confirmed
in Ti-doped samples, where the incommensurate fluctuations
condense to a static spin density wave ordering with the ordered
moment being aligned perpendicular to the planes \cite{srruti}.
The direct analysis in \srruo ~ can be performed by INS. However,
when using unpolarized neutrons the analysis is indirect and
requires some knowledge or assumptions about the underlying form
factor. Our previous rough estimation with INS pointed to some
weak anisotropy \cite{braden2002}, whereas the deeper study by
Servant et al. \cite{servant} concluded that the incommensurate
fluctuations would be isotropic. Another INS study with
unpolarized neutrons concludes a high anisotropy of about a factor
4 \cite{urata}.

In this work, we have used polarized neutrons in order to have
direct access to the anisotropy of the quasi-antiferromagnetic
fluctuations. In addition we have again looked for the existence
of quasi-ferromagnetic fluctuations.

\begin{figure}[tp]
\resizebox{0.93\figwidth}{!}{
\includegraphics*{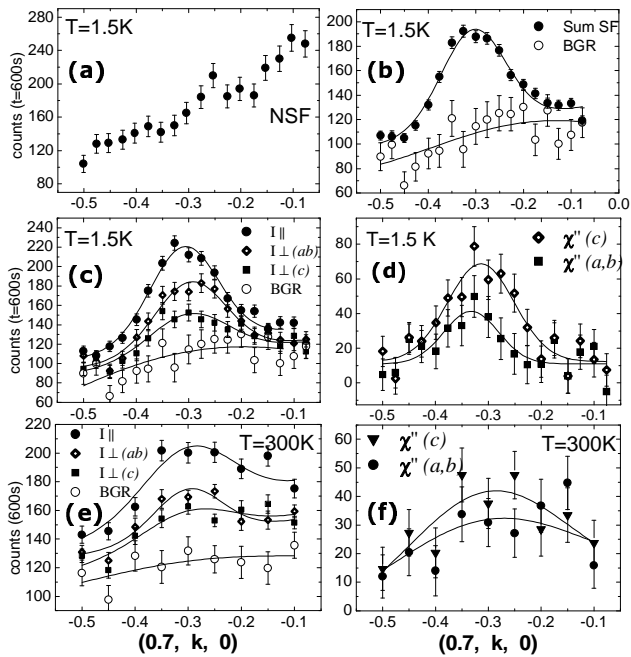}}
\caption{ Results of the scans across the incommensurate position
at \vQ =(0.7,0.3,0) for an energy transfer of 8meV and at T=1.5K
a-d) and at T=300K e-f). Part a) shows the non-spin-flip scan data
which does not exhibit a peak. Part b) shows the total sum of the
spin-flip scattering and the corresponding spin-flip background.
Part c) presents the spin-flip scattering for the different
orientations of the neutron polarization including the spin-flip
background. In part d) we show the out-of-plane and in-plane
susceptibilities obtained from c) and equations (3-5). Parts e)
and f) show the corresponding results obtained at 300K. }
\label{fig1}
\end{figure}

Large single crystals of \srruo ~were grown by a floating zone
technique; the superconducting transition temperatures of all the
samples used were above 1.3 K. Due to their stick like form, ten
crystals of less than 30mm length were coaligned yielding a total
mass of about 10g. INS experiments were performed on the IN20
triple axis spectrometer with polarization analysis on the
monochromator and on the analyser sides as obtained by focusing
Heusler crystals. Scans were performed with a final energy of 35
meV where a pyrolithic graphite filter allows efficient
suppression of the second order contamination. The polarization
ratio, R,  was obtained by measuring phonon and Bragg scattering
to R$\sim $ 15. The monochromator and analyzer crystals polarize
or analyze the neutron spin perpendicular to the scattering plane.
In addition Helmholtz coils were used in order to rotate the
polarization of the neutron at the sample to the desired
orientation. A neutron spin flipper was put in front of the
analyzer crystal. Since there is no incoherent magnetic signal
from nuclear spins in \srruo , the analysis of the spin-flip
signal directly gives the distinct components of the generalized
susceptibility given by equ. (1). We put the sample in the
[100]/[010] orientation and determined the scattering intensities
in the three spin-flip channels with the neutron polarization (i)
parallel to the scattering vector \vQ , $I_\|$, (ii)
perpendicular to \vQ ~ and within the $a,b$-plane, $I_{\bot
a,b}$, and (iii) perpendicular to \vQ  ~and perpendicular to the
planes, $I_{\bot c}$. Since we studied only \vQ -vectors within
the $a,b$-plane, these three measured intensities sense the
following components of the susceptibilities (neglecting
corrections of the order of $1/R$) \cite{shirane}:

\begin{eqnarray}
 I_{\|}       &=& scale \cdot( \chi"_{a,b}  + \chi"_{zz}) +I_{\rm BGR} \, \\
 I_{\bot a,b} &=& scale \cdot  \chi"_{zz}   +              I_{\rm BGR} \, \\
 I_{\bot c}   &=& scale \cdot  \chi"_{a,b}  +      I_{\rm BGR} \,  \label{eq:pol3}
\end{eqnarray}
where $I_{\rm BGR}$ is the spin-flip background scattering. In
addition, for some cases, we have also determined the non-spin
flip scattering to exclude a possible phonon contamination
through the non-ideal polarization ratio. Typically, $I_\|$ was
measured twice as long as the other components and the single
contributions were obtained by subtracting the corresponding
intensities assuming that $scale$ and $I_{\rm BGR}$ are the same
in all spin-flip channels.

Fig. 1 a) and b) show the raw scan data across the incommensurate
position \vQ =(0.7,0.3,0) in the non-spin-flip and spin-flip
channels respectively, the energy transfer was fixed to 8meV at
the temperature of 1.5K. The peak, which is present in the
spin-flip channel only, confirms the magnetic character of the
incommensurate fluctuation, as it was previously deduced from its
\vQ - and temperature dependencies
\cite{sidis,braden2002,servant,servant1}.
In addition there is some weakly $q$-dependent scattering in the
spin-flip channel. Fig. 1c) presents the data in the single
channels (equations (3-5)), and Fig. 1d) the resulting in-plane
and out-of plane components of the susceptibility. We find a
clear anisotropy in disagreement with a previous analysis
\cite{servant,comment}. The out-of-plane susceptibility
$\chi"_{zz} ({\bf q_i},8{\rm meV})$ is about twice as large as
$\chi"_{a,b}({\bf q_i},8meV)$ when considering the peak height
and about 2.5 times as large in the integrated signal. This
anisotropy seems to disappear upon heating to 300K, see Fig. 1 e)
and f).

The sign and the temperature dependence of the magnetic anisotropy
are in qualitative agreement with the calculations, but the
experimental anisotropy at 1.5\ K is much smaller than the
calculation reported in reference \cite{eremin} for comparable
energy transfer. The experimentally determined anisotropy is,
furthermore, slightly smaller than the factor of three obtained
from the analysis of the NMR experiments by Ishida et al.
\cite{ishida2001}. The slight difference  should reflect the much
lower fluctuation energy sensed by the NMR experiment in respect
with the value of 8meV studied here. Eremin et al. indeed
calculate a pronounced energy dependence for the anisotropy ratio
at low temperature suggesting that only the out-of-plane
component is approaching a SDW phase transition. In our previous
work we have analyzed the energy dependence of the magnetic
response at \qi ~ with single relaxor behavior \cite{moriya}:
\begin{equation}
\chi "({\bf q_i}, \omega) = \chi '({\bf q_i}, \omega =0)
\frac{\Gamma \omega}{\omega ^2 +
 \Gamma ^2}
\; \label{eq:3}
\end{equation}
where $\Gamma$ is the characteristic energy and $\chi '({\bf
q_i},0)$ the amplitude which corresponds to the real part of the
generalized susceptibility at \om =0 according to the
Kramers-Kronig relation. In strength, one would need to allow for
two different characteristic energies and amplitudes for the
in-plane and out-of-plane spectra which become superposed in the
unpolarized experiment. Such analysis, however, cannot be made
with the statistics achievable. Fig. 2 shows scans across the
incommensurate position for different energies in the spin-flip
channel with polarization parallel to \vQ . The incommensurate
scattering can be followed up to 40meV whereas the unpolarized
analyzes were restricted to the energy range below 12meV due to
contaminations with phonon scattering. Besides the conclusion that
the nesting signal extends to higher energies than previously
studied, we obtain the anisotropy ratios at higher energies. The
incommensurate signal appears to become more isotropic with
increasing energy. However, the statistics of these data remains
rather poor due to the larger Q-value required for the scattering
geometry and due to the limitation of the complete polarization
analysis to single points. Fig. 3) resumes all results together
with the NMR-anisotropy at low frequency. Taking account of the
average between in- and out-of-plane susceptibility determined in
the unpolarized INS experiments \cite{sidis,braden2002}, one may
describe such energy dependence with $\Gamma _{zz}\sim$8meV,
$\chi '_{zz}({\bf q_i},0)\sim 220\mu _B^2/eV$ and $\Gamma
_{a,b}\sim$13meV, $\chi '_{a,b}({\bf q_i},\omega =0)\sim 140\mu
_B^2/eV$, see the line in Fig. 3. Qualitatively, this behavior
agrees with the analysis by Eremin et al., but the experimentally
observed differences in $\Gamma$ and $\chi '({\bf q_i},\omega
=0)$ are much smaller (this disagreement is far beyond the
statistical errors).

\begin{figure}[tp]
\resizebox{0.5\figwidth}{!}{
\includegraphics*{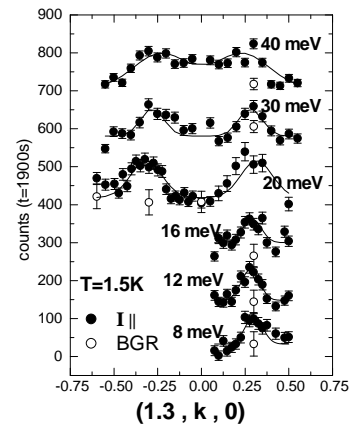}}
\caption{ Results of the scans across the incommensurate position
at \vQ =(1.3,0.3,0) for different energies at T=1.5K. Closed
symbols designate the spin-flip scattering in the
$I_{\|}$-channel and open symbols the spin-flip background. Lines
are guides to the eye. } \label{fig2}
\end{figure}

\begin{figure}[tp]
\resizebox{0.45\figwidth}{!}{
\includegraphics*{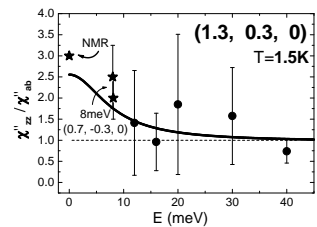}}
\caption{ Energy dependence of the anisotropy ratio
$\chi"_{zz}\over \chi"_{a,b}$ of the incommensurate fluctuations
measured at (1.3,0.3,0) (points); the NMR result by Ishida et al.
\cite{ishida2001} and the two values obtained at \vQ =(0.7,0.3,0)
are denoted by stars, the line represents the energy dependence
obtained for the parameters discussed in the text.} \label{fig3}
\end{figure}

In the theory of Kuwabara and Ogata \cite{kuwabara}  the ability
of the incommensurate fluctuations to stabilize a triplet pairing
is studied quantitatively. Following reference \cite{kuwabara} we
take $S({\bf q=0})$ to 0.8 and use the q-dependence of the
interaction I(q) given in \cite{mazin2}. With the experimental
spin-susceptibility of $\chi '({\bf 0},0)$=0.9$\cdot
10^{-3}$emu/mol$\sim$28$\mu_B^2/{\rm ev}$ \cite{maeno-rev} and the
absolute values of $\chi '_{zz}({\bf q_i},0)$ determined above
\cite{sidis} one obtains the enhancement parameter at the
incommensurate \vq -value : $S({\bf q_i})$=0.97 \cite{comment2}.
The resulting anisotropy in the enhanced susceptibility -- taking
equation (2) for the two channels \cite{kuwabara} -- is then
driven only by a small anisotropy in $\chi^0$ : for an anisotropy
in the enhanced part of 2.5 one gets $\chi'^0_{zz}({\bf
q_i},0)\over \chi'^0_{a,b}({\bf q_i},0)$=1.045, whereas triplet
pairing is stabilized only for ${\chi'^0_{zz}({\bf q_i},0)\over
\chi'^0_{a,b}({\bf q_i},0)}>1.14$  \cite{kuwabara}.  Comparing the
ratios in the enhanced parts one finds that for $S({\bf
q_i})$=0.97 the theory requires an effective anisotropy larger
than 5.5 compared to the experimental value of 2.5. The energy
dependence of the anisotropy discussed above will even enhance the
discrepancy since it reduces the anisotropy in the real part of
the susceptibility at zero energy; the rough estimate amounts
only to a factor of 1.6, see above. Therefore, the sizeable
anisotropy reported here does still not allow to explain triplet
superconductivity within the theory of \cite{kuwabara}; a more
precise theoretical treatment is highly desirable.

\begin{figure}[tp]
\resizebox{0.6\figwidth}{!}{
\includegraphics*{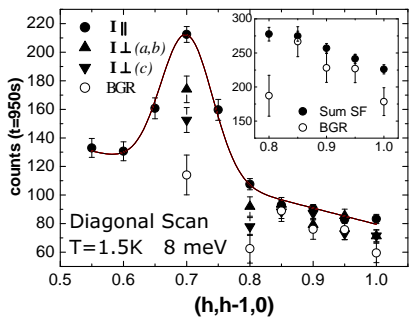}}
\caption{ Results of the scans across the incommensurate position
at (0.7,0.3,0) for 8meV and 1.5K in diagonal direction. For k=1,
\vQ ~ is (1,0,0) corresponding to the two-dimensional
ferromagnetic zone center, all polarization configurations are
shown. The inset presents the spin-flip scattering and its
background close to the ferromagnetic zone-center.}\label{fig4}
\end{figure}

Fig. 4 shows the results of a scan across \vQ =(0.7,0.3,0) ~ in
diagonal direction connecting to the two-dimensional zone-center
(1,0,0) where scattering related to a ferromagnetic instability
should be observable. In agreement with all unpolarized previous
studies \cite{sidis,braden2002,servant} one may conclude that
there is no strong quasi-ferromagnetic scattering for the energy
transfer studied, 8\ meV. The statistics is not sufficient for
complete polarization analysis, but the spin-flip analysis still
indicates some broad magnetic scattering around \vQ =(1,0,0).
Correcting for the form-factor, this signal is about a factor of
five weaker than that of the peak in the incommensurate signal in
rough agreement with the mapping of scattering with unpolarized
neutrons \cite{braden2002} . Furthermore, nearly the same factor
of five can be seen in the scans in Fig. 1b) and d) at k$\sim$0.1
and at low temperature, and this signal seems to increase with
temperature, see Fig. e) and f). Such weakly \vq - and
temperature-dependent scattering perfectly agrees with the
interpretation of the NMR experiments \cite{ishida2001}.

In conclusion INS with polarized neutrons has shown that there is
an anisotropy in the dynamic susceptibility at the incommensurate
position. The enhancement of the out-of-plane component by a
factor of 2-2.5 for an energy transfer of 8meV qualitatively
confirms calculations based on spin-orbit coupling. But the
anisotropy as well as its temperature dependence is much weaker
than the theoretical prediction. Comparing the anisotropy ratio
at 8meV with results at higher energy and with the NMR-analysis
referring to very low energy, one has to conclude some energy
dependence of the anisotropy. Again there is qualitative
agreement with theory, but the experimentally observed energy
dependence is much smaller. Close to the quantum critical point
of the related SDW ordering, the anisotropy, however, should
strongly increase or even diverge as an instabiliy is achieved
only in the out-of-plane channel. An anisotropic analysis will
then be necessary. Similar anisotropic effects should be relevant
in many materials approaching a quantum critical point. Besides
the well studied incommensurate scattering the polarization
analysis gives evidence for additional much weaker scattering
with little \vq -dependence.

{\bf Acknowledgements} We thank I. Eremin and P. Pfeuty for
discussions. This work was supported by the Deutsche
Forschungsgemeinschaft through the Sonderforschungsbereich 608
and by grants from the Ministry of Education, Culture, Sports,
Science and Technology of Japan, and from Japan Society for
Promotion of Science.
* electronic mail : braden@ph2.uni-koeln.de

\end{document}